# Enhancing Optical Readout from Diamond AFM Tips for Quantum Nanosensing

**Sumin Choi[1†], Victor Leong[1,2*†], Gandhi Alagappan[3] and Leonid Krivitsky[1,2]**

[1]Data Storage Institute, Agency for Science, Technology and Research (A*STAR), 2 Fusionopolis Way, #08-01 Innovis, 138634 Singapore

[2]Institute of Materials Research and Engineering, Agency for Science, Technology and Research (A*STAR), 2 Fusionopolis Way, #08-03 Innovis, 138634 Singapore

[3]Institute of High Performance Computing, Agency for Science, Technology and Research (A*STAR), 1 Fusionopolis Way, #16-16 Connexis North, 138632 Singapore

* victor_leong@imre.a-star.edu.sg

[†]These authors contributed equally to this work.

## Abstract

Color centers in diamond are promising candidates for quantum nanosensing applications. The efficient collection of the optical signal is the key to achieving high sensitivity and resolution, but it is limited by the collection optics. Embedding the color centers in diamond microstructures can help to enhance the collection efficiency, but often require challenging fabrication and integration. Here we investigate the photoluminescence (PL) of silicon-vacancy (SiV) centers in commercially available atomic force microscope (AFM) diamond pyramid (DP) tips. We find that the DP geometry efficiently channels PL emitted at the DP apex towards the base, where we experimentally demonstrate an enhanced PL collection of up to 8 times higher compared to other directions. Our experimental observations are in good agreement with numerical simulations using a finite-difference time-domain (FDTD) method. Our results indicate that AFM tips could be an economical, efficient and straightforward way of implementing color-center-based nanosensing as they provide enhanced sensitivity and easy integration with existing AFM platforms.

**Introduction**

Color centers in diamond are atomic-sized quantum emitters that could be well-suited for numerous sensing applications requiring nanoscale resolution[1-5]. They are photostable, robust at both cryogenic[6] and at high temperatures[7], and are compatible with biological environments[8]. In particular, the nitrogen-vacancy (NV) center has been widely utilized for precise measurements of electromagnetic fields[9,10], strain[4,11] and temperature[3,12]. Magnetometry with NV centers has led to recent research advanced in superconductivity[13,14], magnetic materials[15,16], and magnetic resonance imaging (MRI) of living cells[17]. Recently, silicon-vacancy (SiV) and germanium-vacancy (GeV) centers have also emerged as viable options for all-optical thermometry[18,19].

To utilize these color centers as practical nanoprobes, recent research has investigated their use in nanodiamonds or nanophotonic structures such as nanopillars[20]. Recent work has also explored commercially available diamond-based atomic force microscope (AFM) tips[21,22], which are robust and relatively low-cost. These tips consist of single-crystal diamond pyramids (DPs) mounted on a cantilever. Given the very small tip radii of ~ 10 nm, emitters near the DP apex can be brought into very close proximity with the sample, which is crucial in maximizing measurement sensitivity and spatial resolution.

Yet, there is another potential advantage of diamond AFM tips that is still largely unexplored. The DP geometry can also be exploited to enhance the fluorescence collection of emitted photoluminescence (PL); considering that the tapered shape of the DP is akin to the waveguide tapers commonly used in silicon photonics[23], one might expect that PL from emitters near the apex would also be efficiently channeled along the DP towards the base. However, this was not investigated in the previous works; as those DP were mounted on metal-coated silicon cantilevers, which are not transparent at visible wavelengths, it was not possible to collect the PL through the pyramid base. Ref [20] did observe some enhancement due to photonic effects, but this was observed only for emission through the DP tip (opposite direction from the base). The lack of optical access to the DP base would, in turn, require more complicated setups to collect the PL signal while maintaining the probe-sample proximity, e.g. at steep angles[24-28] or via transmission, which requires a transparent sample and/or substrate[29-31].

Here we study AFM tips with DPs mounted on transparent silicon nitride (SiN) cantilevers, and measure the PL emitted from the DP base (through the cantilever), as well as that along other different directions with respect to the DP. We then quantify the enhancement, or lack thereof, of PL collected along these directions, and compare our results with Finite Difference Time Domain (FDTD) simulations. The ability to enhance single-photon-level PL signals from emitters at the DP apex leads to the improved sensitivity of these nanoprobes. Moreover, collecting PL through the cantilever simplifies experimental setup geometries, enabling straightforward integration of with conventional AFM or confocal microscopy systems, or even miniaturized systems with photodetectors attached directly to the back of the cantilever.

## Experimental Setup

*Diamond pyramid*

We use commercially available AFM tips (Artech Carbon) with single-crystal DPs grown using a microwave plasma enhanced chemical vapor deposition (CVD) process; some of the fabrication details are described elsewhere[21,32]. Notably, these DPs were grown in nominally nitrogen-free process, resulting in an absence of NV centers, while incorporation of silicon from the substrate during CVD growth led to an abundance of SiV centers throughout the DP. The DPs have a ~5×5 μm$^2$ base, ~14 μm length, and are glued to a SiN cantilever (NanoWorld AG, PNP-TR-TLx00-50) without metal coating. The SiN cantilever is transparent to the SiV zero-phonon line (ZPL) emission at 737 nm[33].

*Confocal microscope*

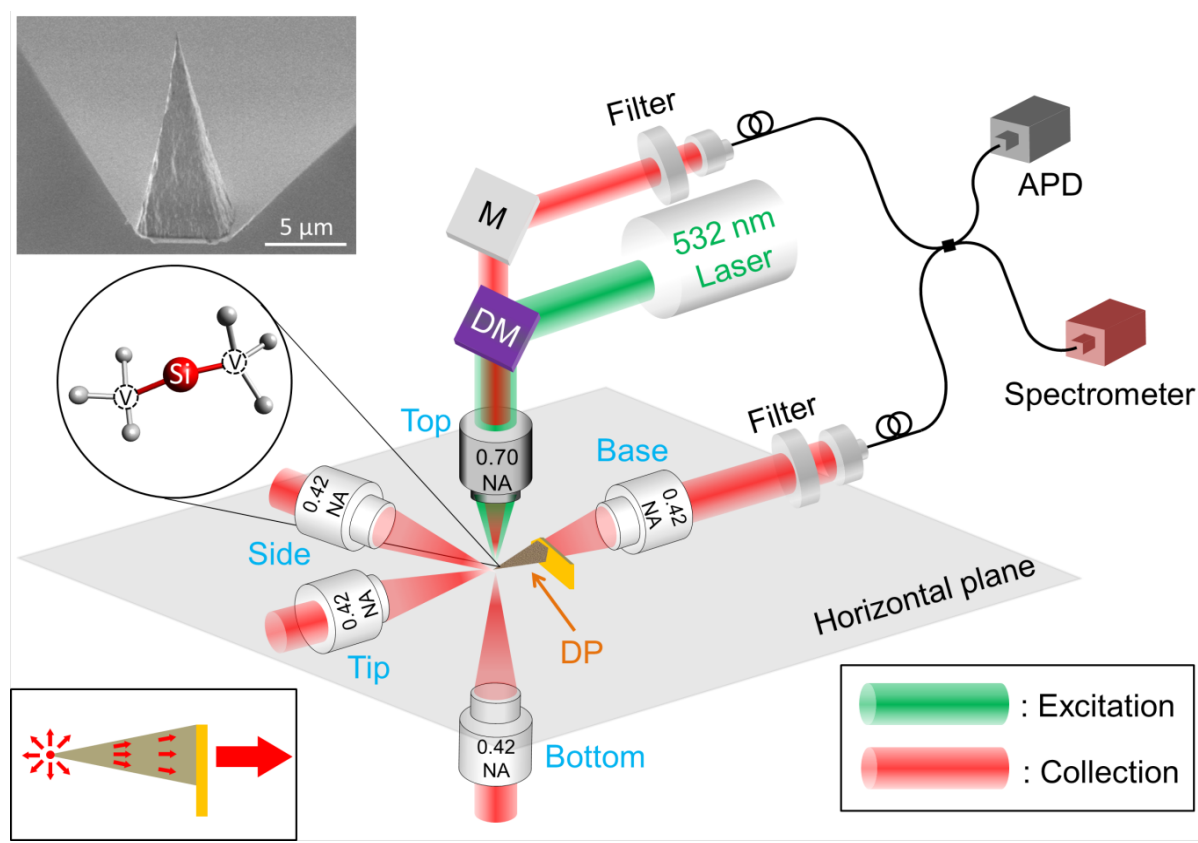

Figure 1. Confocal microscope. A continuous-wave 532 nm excitation laser is focused onto the diamond pyramid (DP). PL is detected along 5 different orientations; in each case, the signal passes through a notch filter and a narrowband filter, is collected into a multi-mode fiber, and sent to an avalanche photodiode (APD) or spectrometer. Insets: (top left) Scanning electron microscope (SEM) image of a DP, (middle left) Sketch of the silicon vacancy (SiV)

atomic structure, (bottom left) Sketch of the channeling effect though the DP. M: mirror, DM: dichroic mirror.

The optical properties of the DPs are studied with a home-built confocal microscope at room temperature (see Fig. 1). The DP is mounted on a 3D piezoelectric stage (Piezosystem Jena, TRITOR 100 SG) with the DP axis along the horizontal plane. Excitation light from a continuous-wave 532 nm laser (Oxxius) is focused onto the DP through an air objective lens (excitation objective, Mitutoyo, NA=0.70, working distance (wd) = 6 mm) with a power of 100 μW. SiV PL is collected both through the excitation objective and a separate objective lens (collection objective, Mitutoyo, NA=0.42, wd = 20.5 mm) that is orientations (tip, base, side, and bottom) with respect to the DP. We choose a smaller NA for the collection objective due to the necessity of mounting two objectives in close proximity for the test measurements. The emission passes through a notch filter (Semrock NF03-532E) and a narrowband filter (Semrock FF01-740/13); these filters reject the excitation light and transmit only near the ZPL. The PL signal is then collected into a multi-mode fiber (Thorlabs GIF625, 62.5 μm core) and directed to either a grating spectrometer (Princeton Instruments IsoPlane 160, 0.07 nm resolution), or to an avalanche photodetector (APD, Perkin Elmer SPCM-AQRH-15) for measuring photon counts.

To perform confocal scans, we first overlap the foci of both objectives at the apex region of the DP, and then move the DP in 100 nm steps in the horizontal plane. For practical reasons, we do not construct separate optical paths for the different orientations in the horizontal plane measured with the collection objective; instead we rotate the DP between each measurement to achieve the same effect without moving the collection objective.

**Experimental Results and Discussion**

We studied 7 DPs in detail, and report our results below.

*Confocal scan images*

Confocal scan images of a DP (sample 1) recorded at different orientations are presented in Fig. 2(a)-(e). The 'Top' image shows the DP outline with the brightest regions in the lower half (towards the base), where there is a larger diamond volume and thus more SiV centers within the confocal volume. We note here that at certain regions of the DP, especially in the upper half, the diamond volume is not sufficiently large to fill the confocal volume, given the modest NA of our objective lenses.

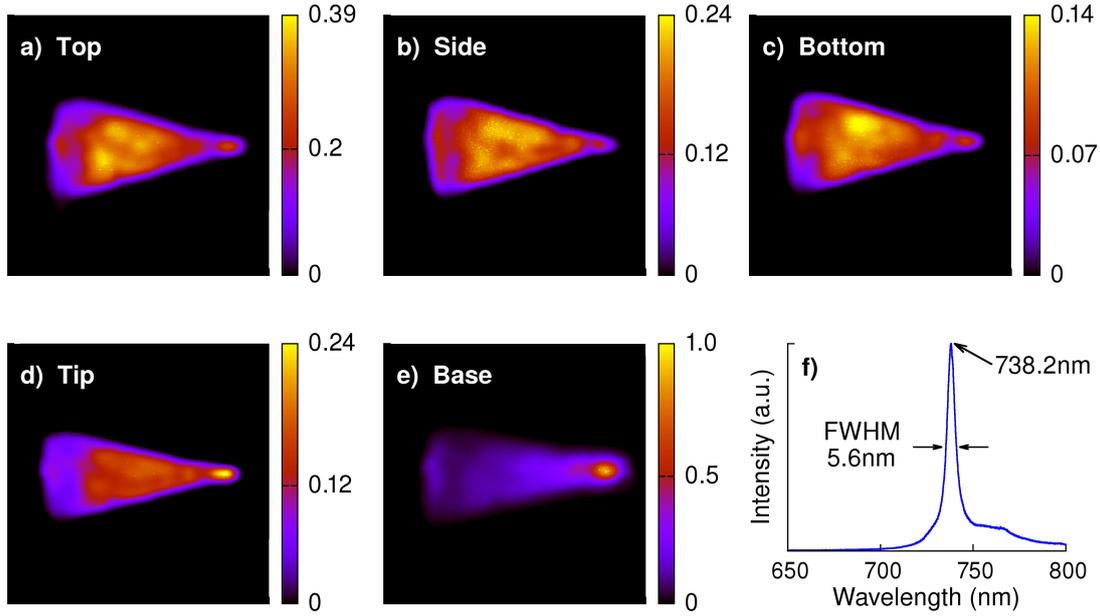

Figure 2. (a)-(e) Confocal scan images obtained from different measurement orientations of a DP (sample 1). Scan area is 20 × 20 μm and pixel size is 100 × 100 nm. Color scale is the same across the figures and represents photon count rates, normalized to the maximum rate of 13.7 × $10^6$ $s^{-1}$ recorded through the base. The maximum value of the color scale of each plot represents the count rate at the brightest pixel within that plot. (f) Photoluminescence (PL) spectrum of the DP, showing the SiV zero phonon line (ZPL) peak at 738.2 nm with a full-width-half-maximum (FWHM) of 5.6 nm, and the phonon side band. There is no observed difference in spectral properties when varying DP position or measurement orientation.

The 'Side' and 'Bottom' images are similar to the 'Top' image, but are dimmer due to the collection objective having a lower NA. The 'Tip' image shows a concentration of higher photon rates in the apex region, despite the smaller number of emitters there. This agrees with previous observations, which have been attributed to some photonic effects, as well as a higher SiV concentration near the apex due to the nature of the CVD growth process[21]. Nonetheless, the count rates in these images are all relatively low.

With the 'Base' orientation, we observe a significant enhancement of photon rates at the apex region, indicating a strong channeling of SiV PL along the DP towards the base. The apex PL collected from the base also exceeds the brightest regions in the 'Top' orientation, despite the higher NA objective used for the latter. To quantify this effect, we find the brightest pixel in the 'Base' image with a photon count rate $R_{base}$, and define an enhancement factor $\alpha_X = R_{base}/R_X$ where $R_X$ is the photon rate at the same location on the DP measured along a different direction X; the uncertainty is then given by our ability to align the two scan images (accuracy within 1 pixel). For sample 1, we obtain $\alpha_{tip}$= 4.29 ± 0.08 and $\alpha_{side}$= 7.75 ± 0.12. The corresponding values for all DPs are shown in Table 1. The variation in the measured enhancement factors is likely due to sample imperfections: scanning electron microscope (SEM) inspections reveal slight variations in the DP shape and surface quality between different samples. We expect that the enhanced PL collection is a broadband effect, and thus our results are promising for developing efficient nanoprobing schemes based on various color centers in DPs.

| Sample no. | Enhancement factor | |
|---|---|---|
| | $\alpha_{tip}$ | $\alpha_{side}$ |
| 1 | 4.29 ± 0.08 | 7.75 ± 0.12 |
| 2 | 2.96 ± 0.11 | 4.13 ± 0.15 |
| 3 | 4.30 ± 0.07 | 8.20 ± 0.20 |
| 4 | 2.52 ± 0.04 | 3.77 ± 0.12 |
| 5 | 2.44 ± 0.04 | 4.62 ± 0.21 |
| 6 | 3.72 ± 0.06 | 8.13 ± 0.09 |
| 7 | 2.85 ± 0.05 | 6.39 ± 0.14 |

Table 1. Enhancement factors ($\alpha_{tip}$ and $\alpha_{side}$) of photon rates measured from the DP base compared to the other orientations.

*PL spectra*

A typical PL spectrum of the DP is shown in Fig. 2(f). It clearly shows the characteristic SiV ZPL at 738 nm with a full-width-half-maximum (FWHM) of 5.6 nm, and a distinct phonon side band (PSB) above 750 nm. Notably we do not observe spectral signatures of NV centers or other defects, indicating a high purity of SiV centers in the DP. Besides the intensity, we do not observe any significant difference in the spectral properties when varying DP position or measurement orientation, or between different DP samples.

*FDTD Simulations*

To better understand the enhancement effect, we aim to compare our experimental results with numerical simulations. We model the DP as a pyramid of refractive index of n = 2.4 with the same dimensions as reported above, and solve Maxwell's equations using a 3D FDTD method[34]. To obtain a representation of the emission pattern in different regions of the DP, we simulate a 100 nm diameter spherical cloud of dipoles emitting at 738nm, and vary its position along the center line of the DP. The cloud consists of 100 dipoles with random positions and orientations (see Fig. 3(a)). We measure the emitted power through three different monitors - side, tip, and base. In order to compare with experimental observations, the powers from these monitors were subjected to far-field filters with NA = 0.42 (same as the collection objective used in the experiment). The simulation results are shown in Fig. 3(b). The power at the base monitor is significantly larger than at the side and tip when the dipoles are at the DP apex, indicating an efficient channeling of emitted light along the DP towards the base, similar to our experimental observations.

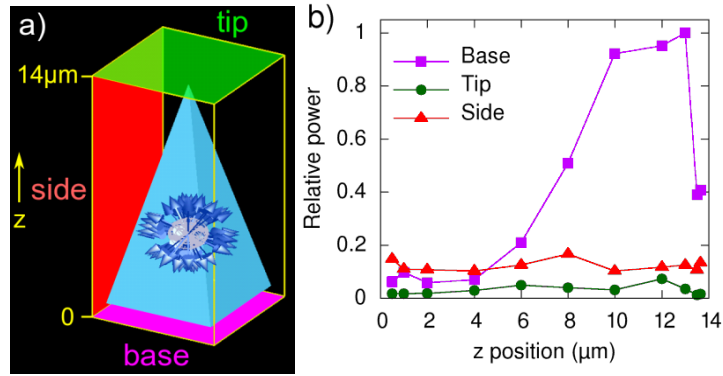

Figure 3. (a) Sketch of the DP in the FDTD simulations, showing the dipole cloud and the orientation of the three monitors. (b) Simulated power at the base, tip, and side monitors for varying positions of the dipole cloud along the center line of the DP.

**Comparing simulations and measurements**

Both measurements and simulations exhibit enhanced PL collection at the DP base. We can compare the two by extracting the intensity profile along the central line of the DP from the confocal scan images, then matching it against the FDTD results. However, a direct comparison is inappropriate: the simulations assume a constant number of emitters at each position, while the effective number of emitters at each pixel varies across the measured confocal images as both the density of SiV centers and the volume of the DP within the confocal volume are not constant. As such, we correct for this by using the 'Bottom' image as an approximation of the effective number of SiV centers within each pixel. To do so, we multiply the simulated curves in Fig. 3b by the center line profile extracted from the 'Bottom' confocal image (Fig. 2c), and then match the corrected curves against the corresponding profiles from the other images (see Fig. 4).

The corrected curves have a good qualitative agreement with our measured results, in particular the sharp peak near the apex when measuring at the base (Fig. 4d). The simulated enhancement factors, which are unaffected by the correction procedure, are $\alpha_{tip}$= 29.4 and $\alpha_{side}$= 8.1. These are much larger than our measured values, with the discrepancy likely due to imperfections in the DP samples. Nonetheless, the general good agreement between simulations and measured results, despite the approximate nature of our correction, indicate that we have adequately understood the channeling effect.

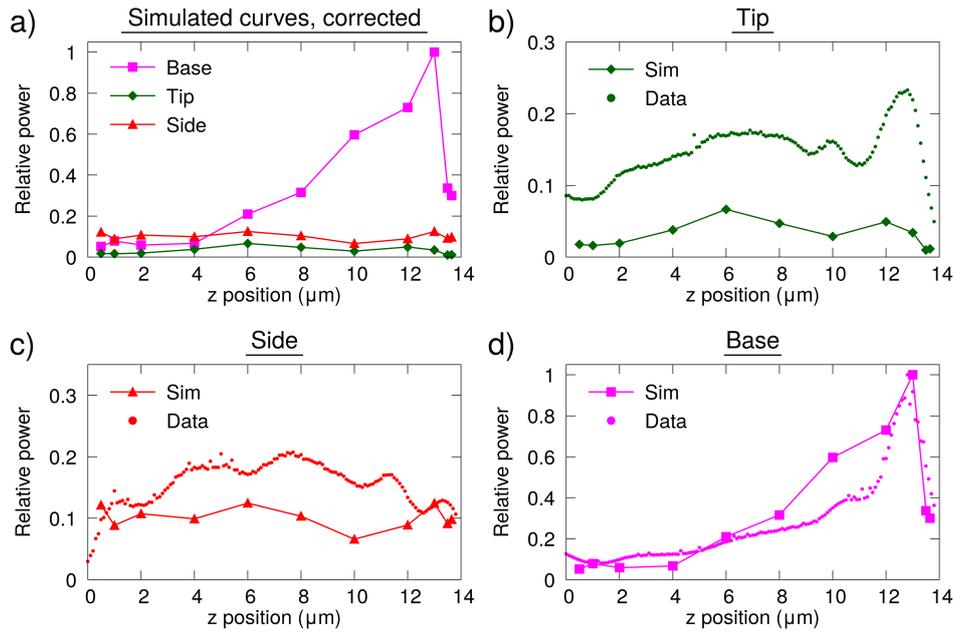

Figure 4. (a) Simulated PL power at the base, tip, and side monitors for varying positions of the dipole cloud along the center line of the DP, corrected using the line profile extracted from Fig. 2c. (b-d) Comparison of PL power between the corrected simulation curves and measured data from sample 1 from the tip, side and base along the DP center line profile. Both the simulated and measured values are normalized to their respective peak value in (d).

**Conclusion**

In conclusion, we have investigated SiV PL from AFM tips consisting of DPs mounted on transparent SiN cantilevers. We measured the PL along different directions of the DPs, and found a strong channeling of PL emitted near the apex towards the DP base. Comparing the photon rates at the apex region measured at the base with that measured along other directions, we find enhancement factors of approximately 2.5–8. There is good qualitative agreement between our experimental observations and FDTD simulations. The PL collection enhancement can be further improved by optimizing the geometry and growth quality of the DPs. Our results show that color centers in diamond AFM tips can be a relatively cheap, straightforward, and efficient approach towards quantum nanosensing applications. For example, a conventional AFM system can be upgraded by integrating a confocal microscope with its optical inspection system, thus extending the functionality of the system into the quantum sensing domain.

**Funding**

This work was supported by NRF-CRP14-2014-04, "Engineering of a Scalable Photonics Platform for Quantum Enabled Technologies".